\documentclass[aps,prb,twocolumn,showpacs,superscriptaddress,10p,longbibliography]{revtex4-2}
\usepackage{amssymb}
\usepackage{graphicx}
\usepackage[colorlinks = true,linkcolor = red,citecolor = blue,urlcolor=blue]{hyperref}
\usepackage[sort&compress]{natbib}
\usepackage{scalerel}
\usepackage[normalem]{ulem}
\usepackage[dvipsnames]{xcolor}
\usepackage{bm}
\usepackage{amsmath}
\usepackage{subfigure}
\usepackage{amsfonts}
\usepackage{mathtools}
\usepackage{dsfont}
\usepackage{float}
\usepackage{lmodern}
\usepackage[many]{tcolorbox}
\usepackage{empheq}
\usepackage{soul}
\usepackage[T1]{fontenc}

\newcommand{\ket}[1]    {|#1 \rangle}

\newcommand{\hs}{\hat{\sigma}}

\newcommand{\up}{\uparrow}
\newcommand{\dn}{\downarrow}

\usepackage{tikz}
\definecolor{lime}{HTML}{A6CE39}
\DeclareRobustCommand{\orcidicon}{%
    \begin{tikzpicture}
    \draw[lime, fill=lime] (0,0) 
    circle [radius=0.13] 
    node[white] {{\fontfamily{qag}\selectfont\tiny ID}};
    \draw[white, fill=white] (-0.0625,0.095) 
    circle [radius=0.007];
    \end{tikzpicture}
    \hspace{-2mm}}
    
\newcommand{\orcidAM}{\href{https://orcid.org/0000-0002-7307-5922}{\orcidicon}}
\newcommand{\orcidJK}{\href{https://orcid.org/0000-0003-0998-9460}{\orcidicon}}

\newcommand{\orcidMP}{\href{https://orcid.org/0000-0002-0835-1644}{\orcidicon}}

\newcommand{\orcidML}{\href{https://orcid.org/0000-0002-0210-7800}{\orcidicon}}

\begin{document}

\title{String breaking dynamics in
Ising chain with local vibrations}

\author{Arindam Mallick\orcidAM}
\affiliation{Instytut Fizyki Teoretycznej, Wydzia\l{} Fizyki, Astronomii i Informatyki Stosowanej, Uniwersytet Jagiello\'nski, \L{}ojasiewicza 11, PL-30-348 Krak\'ow, Poland}

\author{Maciej Lewenstein\orcidML}
\affiliation{ICFO-Institut de Ciencies Fotoniques, The Barcelona Institute of Science and Technology, 08860 Castelldefels (Barcelona), Spain}
\affiliation{ICREA, Passeig Lluis Companys 23, 08010 Barcelona, Spain}

\author{Jakub Zakrzewski\orcidJK}
\affiliation{Instytut Fizyki Teoretycznej, Wydzia\l{} Fizyki, Astronomii i Informatyki Stosowanej, Uniwersytet Jagiello\'nski, \L{}ojasiewicza 11, PL-30-348 Krak\'ow, Poland} 
\affiliation{Mark Kac Complex Systems Research Center, Jagiellonian University in Krak\'ow, PL-30-348 Krak\'ow, Poland}

\author{Marcin Płodzień\orcidMP}
\affiliation{ICFO-Institut de Ciencies Fotoniques, The Barcelona Institute of Science and Technology, 08860 Castelldefels (Barcelona), Spain}

\date{\today}

\begin{abstract}

We consider the dynamics in the one-dimensional quantum Ising model in which
each spin coherently interacts with its phononic mode. The model is motivated by quantum simulators based on Rydberg atoms in tweezers or trapped ions. The configuration of two domain walls {simulates} the {particle-antiparticle connecting} string. We concentrate on the effect the local vibrations have
on the dynamics of this initial state. Our study supplements recent investigations of string breaking, traditionally studied within quantum chromodynamics (QCD), to quantum many-body systems. Two regimes are identified depending on the strength of the coupling with local vibrations. For weak coupling, the string breaking is slowed down as compared to the dynamics in an isolated Ising string. The strong coupling leads to complicated dynamics in which the domain wall character of excitation is dissolved among many coupled states.
\end{abstract}

\maketitle
\nopagebreak

\section{Introduction}
 
Quantum chromodynamics (QCD) is a non-Abelian gauge theory formulated on the SU(3) symmetry group, describing the strong interaction, which is a fundamental force responsible for the binding of quarks and gluons into protons, neutrons, and other hadrons \cite{GellMann1964, Marciano1979,RevModPhys.82.1349}. The theory's complexity is attributed to features such as asymptotic freedom \cite{PhysRevLett.30.1343,PhysRevLett.30.1346} and color confinement \cite{POLYAKOV1977429,PhysRevD.10.2445,THOOFT19781,ALLAN1977375}, which lead to a variety of phenomena observable in high-energy particle collisions and the detailed structure of atomic nuclei.
In particular, quark confinement occurs due to the squeezing of the chromoelectric flux into a string-like structure known as the QCD string, a consequence of nonperturbative vacuum effects \cite{PhysRevLett.81.4056, KNECHTLI2000673,DmitriAntonov_2003, PhysRevD.84.065013, Greensite2011}. 
The string breaks beyond a critical separation distance due to the creation of quark-antiquark pairs.

\begin{figure}[h]
\includegraphics[width = 0.98\columnwidth]{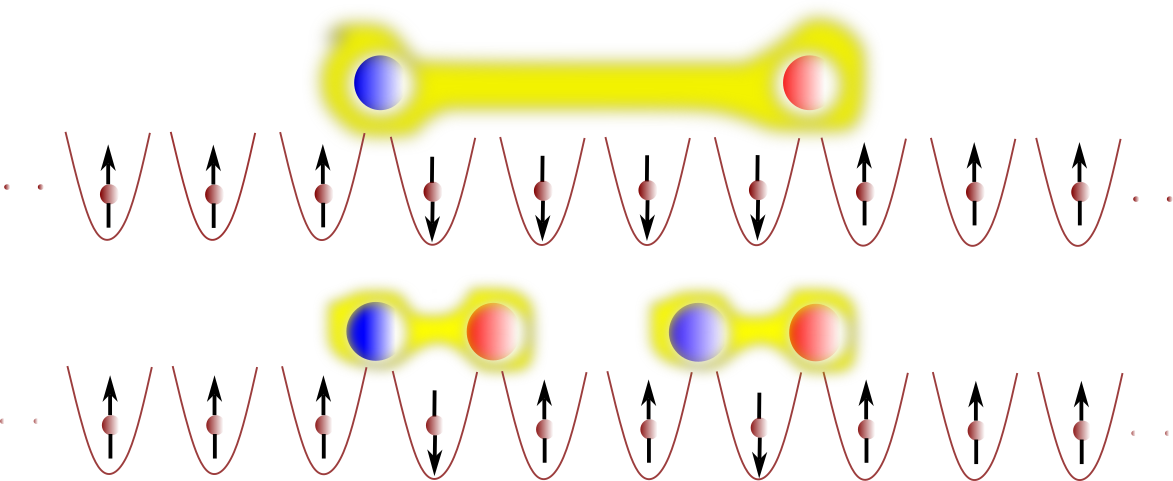}
\caption{Top: Exemplary initial state as defined in Eq.~\eqref{eq:initial_state}. Brown circles indicate the mean positions of lattice sites that are trapped inside quadratic wells (indicated by parabolas), and the sites can vibrate like quantum harmonic oscillators with respect to their mean position. Each site carries $\up$ or $\dn$ spin. Two-domain walls are associated with a particle-antiparticle pair shown in blue and red circles, while the pair connecting string (analogous to electric gauge-flux) is shown as a yellow-shaded ribbon. Bottom: String-broken state: the creation of additional two domain walls inside the initial string.}
\label{fig:initial_state}
\end{figure}

One of the most important avenues in high-energy physics is the real-time dynamics of QCD phenomena. 
In the last few years the studies of Lattice Gauge Theory (LGT) models have been tackled from the point of view of quantum simulators \cite{banerjee2012atomic,WIESE2014246,banuls2020simulating, Aidelsburger2021, PhysRevResearch.5.033184, PhysRevD.108.094513, ArgelloLuengo2024}. Quantum Simulators are anticipated to have the capability to directly investigate the real-time dynamics of quantum field theories \cite{Halimeh2020, Halimeh2022, Halimeh2022_2, Osborne2023,  Halimeh2023, Herschel2023, Popov2024, Zhang2024, Calaj2024,feldmeier2024,Ciavarella2024,ciavarella2024quantumsimulationlargen,grabowska2024}. Recent advancements in quantum hardware have prompted research into the implementation of LGT simulations on quantum computers. The initial quantum simulations of pure non-Abelian LGT's have been conducted in low dimensions using quantum hardware \cite{martinez2016real,nguyen2022digital,farrell2024scalable}.

In recent years, the problem of string breaking dynamics has been studied from the quantum simulator perspective in the paradigmatic quantum Ising chains \cite{markus2020realtime, PhysRevLett.109.175302, PhysRevLett.131.230402, PhysRevD.108.094518,Surace24adia,bacciconi2025}. In this context, the elementary excitations are domain walls that can experience confining potentials due to either symmetry-breaking fields or long-range interactions. The gauge flux or string that connects elementary excitations in LGT corresponds to the magnetic domain of spin-down states between the domain walls in the quantum Ising chain. This broadens the concept of string breaking to quantum many-body theory and makes it accessible to experimental studies using quantum simulators, such as Rydberg atoms \cite{Arinjoy2024,PhysRevX.10.021041,gonzalezcuadra2024}, trapped ions \cite{Crippa2024}, superconducting qubits \cite{ciavarella2024stringbreakingheavyquark} or optical lattices \cite{Liu24exp}. This could help to solve the existing puzzle of nonperturbative QCD.

In this work, we examine an unconventional—but experimentally viable—scenario from the perspective of  LGT that is nonetheless well suited to quantum-simulator platforms. Specifically, we study a model in which (anti-) particles couple to local vibrational modes—a construction that may seem ad hoc in LGT but arises naturally in many quantum-hardware architectures. In these platforms, the spatial “location” of each trapped particle (for example, an atom or ion whose internal states encode a spin-$1/2$ degree of freedom) is described by the expectation value and width of its motional wave packet. Because particles in quantum simulators interact via long-range electromagnetic forces, the dependence of their interaction energy on their positions can be recast in terms of effective phonon degrees of freedom, which has been implemented in analog quantum simulator platforms based on long-range-interacting ultracold atoms or ions  \cite{Wster2011,Hague2012,Hague2013,Barredo2015,Plodzien2018,Camargo2018,Jachymski2020,Mendona2023,Magoni2022,DiLiberto2022,Kosior2023,Magoni2023,Knorzer2022,Magoni2024}.

Our work focuses on string-breaking dynamics in the presence of phonons. Namely, we study the domain wall dynamics in the one-dimensional quantum Ising model where each spin interacts with local vibrations, i.e., non-dispersive phonons.  We show numerically that the string-breaking mechanism is 
suppressed
due to the energy being exchanged with phonons i.e., excitations of the vibrational fields.
 
The manuscript is organized as follows: in Section~\ref{sec:Model} we introduce the system Hamiltonian.  {  In Section \ref{sec:Polaron_picture}, we apply a polaron transformation to the model, replace explicit spin–phonon coupling with spin-dressed operators, and then analyze the resulting effective Hamiltonian semi-classically using coherent phonon states.}. Next, in Section~\ref{sec:Observables} we introduce studied initial state of the string, and set of observables used to characterize string-breaking dynamics. In Section~\ref{sec:Results} we present { {many-body}} numerical findings for the string-breaking dynamics in the short-range interacting one-dimensional Ising models. We discuss the results and conclude in Section~\ref{sec:Conclusions}.

\section{Model }\label{sec:Model}

We consider the one-dimensional spin-$1/2$ chain of length $L$ interacting with dispersionless local vibrations, {  namely the Ising-Holstein model described by the Hamiltonian}
\begin{equation}\label{eq:H_model}
    \hat{H} = \hat{H}_{0} + \hat{H}_{\rm ph} + \hat{H}_{\rm int},
\end{equation}
with 
\begin{equation}\label{eq:H_0}
\begin{split}
 \hat{H}_{0} &= -\sum_{j = 1}^{L-1}\hs^z_j\hs^z_{j+1} - h^x\sum_{j = 1}^{L}\hs^x_j - h^z\sum_{j = 1}^L\hs^z_j,\\
 \hat{H}_{\rm ph} & = \omega_0\sum_j\hat{a}^\dagger_j\hat{a}_j ,\\
 \hat{H}_{\rm int} & = g\sum_{j=1}^{L}(\hat{a}^\dagger_j + \hat{a}_j)\hat{\sigma}^z_j,
\end{split}
\end{equation}
where $\hat{\sigma}_i^{\beta}$, $\beta = x,y,z$ are Pauli operators while $\hat{a}_j (\hat{a}^\dagger_j)$ are bosonic annihilation (creation) operators representing local vibrations, fulfilling bosonic commutation relations $[\hat{a}_j, \hat{a}^\dagger_i]=\delta_{i,j}$. The term 
$\hat{H}_0$ describes the quantum Ising model with the longitudinal and transverse magnetic fields (with amplitudes $h^z$, and $h^x$, respectively) in the open boundary conditions geometry, while the spin-spin interaction is short-ranged. The $\hat{H}_{\rm ph}$ is a local vibrational Hamiltonian with energy scales set by $\omega_0$ which controls the depth of the trapping potential around each lattice site,  see Fig.~\ref{fig:initial_state}. 
 $\hat{H}_{\rm int}$ is the spin-phonon interaction Hamiltonian, where each $z$-spin component interacts with local vibrations with amplitude $g$. Larger displacement $\equiv (\hat{a}^\dagger_j + \hat{a}_j)$ of vibrating lattice site $j$ induces larger interaction energy for a fixed interaction strength $g$, which in turn modifies the effect of the longitudinal magnetic field $h^z$. $\hat{H}_{\rm ph} + \hat{H}_{\rm int}$ resembles a Hamiltonian of displaced quantum harmonic oscillator where the displacement depends on the orientation of $z$-spin. 
 
 According to the mapping between $\mathbb{Z}_2$ LGT and quantum Ising chain, the mass of the matter particle is related to the ZZ-interaction strength, $h^x$ controls the gauge-field induced interaction between particles/antiparticles and responsible for particle creation and annihilation, finally $h^z$ is the string-tension strength which controls the energy of the gauge flux \cite{surace2021phdthesis}.  

{ 

\section{Spin-dressed picture}\label{sec:Polaron_picture}

The Ising-Holstein model can be transformed to an effective Hamiltonian when spin-phonon coupling term is removed, by the cost of spin-dressed operators. In the following we perform the polaron transformation of the Ising-Holstein model, and  discuss the effective semi-classical description assuming coherent states of phonons.

\subsection{Polaron Transformation}

We start with the canonical Lang-Firsov (LF) transformation \cite{LangFirsov1963} to remove the linear spin-phonon interaction term. We consider the unitary transformation $\hat{U} = e^{-\hat{S}}$
\begin{equation}
\hat{S} = \gamma\sum_i (a_i^\dagger - a_i)\sigma_i^z,
\end{equation}
where  $\gamma = -g/\omega_0$   under which the Ising-Holstein model reads $ \hat{\cal H}_{\rm LF} = \hat{U}\hat{H}\hat{U}^\dagger$, 
\begin{equation}
\begin{split}
    \hat{\cal H}_{\rm LF}  =  &-\sum_j \hat\sigma_j^z \hat\sigma_{j+1}^z - h^z\sum_j \hat{\sigma}^z_j + \omega_0\sum_j \hat a_j^\dagger \hat a_j \\
    & - h^x\sum_j (\hat\sigma_j^+ e^{- 2\gamma(\hat{a}^\dagger_j-\hat{a}_j)} + \hat\sigma_j^- e^{2\gamma(\hat{a}^\dagger_j-\hat{a}_j)}) - \frac{g^2}{\omega_0},
    \end{split}
\end{equation}
where $-g^2/\omega_0$ is a constant energy shift. 
The LF transformation results in spin-phonon dressing of the $\hat{\sigma}^x_j$ operators. Assuming a vacuum for the phonons, the transverse field is renormalized as $h^x \mapsto e^{-2\gamma^2} h^x$. 
As 
such,  the effect of spin-phonon coupling on the spin dynamics can be interpreted as suppressing the spin-flip processes governed by the $\hat{\sigma}^x_j$ terms.

\subsection{Semi-classical description}

To gain insight into the role of phonons on the spin dynamics, we start by analysing the Ising-Holstein model in the semi-classical description assuming phonons are in a coherent state.

To obtain the equation of motion for phonons, we apply the Dirac-Frenkel-McLachlan time-dependent variational principle \cite{Dirac1933,Frenkel1934,McLachlan1964}, based 
 on the idea of minimizing an action functional 
 \begin{equation}
\mathcal{S}[\Psi] = \int dt\,   {\cal L} \ ,
\end{equation}
for a time-dependent quantum state, with the Lagrangian ${\cal L} = \langle\Psi(t)|i\partial_t - \hat{\cal{H}}_{\rm LF}|
\Psi(t)\rangle$. The variational principle states that the evolution of $\ket{\Psi(t)}$ should make the action stationary,
$\delta \mathcal{S} = 0$,
under variations $\delta \ket{\Psi(t)}$ \cite{Zhao1997,Raab2000,Perroni2004,Wu2013,Yuan2019}. 

We consider a Davydov-like  time-dependent variational ansatz \cite{Davydov1977}:
\begin{equation}
\ket{\Psi(t)} = \ket{\psi(t)} \otimes \prod_j \ket{\alpha_j(t)}\equiv \ket{\psi(t)}\ket{\{\alpha_j(t)\}},
\end{equation}
where $\ket{\psi(t)}$ is quantum state for spins, while $\ket{\alpha_j(t)}$ are phonon coherent states at each lattice site. The Lagrangian reads  
\begin{equation}
\mathcal{L}
 =\langle\psi|i\partial_t|\psi\rangle
  +\sum_{j}\frac{i}{2}\bigl(\alpha_j^{*}\dot{\alpha}_j-
                            \dot{\alpha}_j^{*}\alpha_j\bigr)
  -\langle\psi|\hat{{\cal H}}_{\rm eff}[\{\alpha_j(t)\}]|\psi\rangle,
\end{equation}
where
$ \hat{{\cal H}}_{\mathrm{eff}}\!\bigl[\{\alpha_j(t)\}\bigr]
  = \bigl\langle\{\alpha_j(t)\}\bigr|\hat{H}_{\mathrm{LF}}
    \bigl|\{\alpha_j(t)\}\bigr\rangle $.
The set of classical-quantum equations of motion describing the evolution of the quantum spins coupled to phonons in coherent states reads:
\begin{equation}\label{eq:EOM}
    \begin{split}
i \dot{\alpha}_j & =  \omega_0 \alpha_j  + 2\gamma h^xe^{-2\gamma^2} \left( \langle \hat{\sigma}_j^+ \rangle e^{-i\theta_j} - \langle \hat{\sigma}_j^- \rangle e^{i\theta_j} \right),\\
i|\dot{\psi}(t)\rangle & = \hat{{\cal H}}_{\rm eff}[\{\alpha_j(t)\}]|\psi(t)\rangle,\\
\hat{{\cal H}}_{\rm eff}[\{\alpha_j(t)\}] &= -\sum_{j} \hat{\sigma}_j^z \hat{\sigma}_{j+1}^z - h^z \sum_j \hat{\sigma}_j^z + \omega_0 \sum_j |\alpha_j|^2\\
&- e^{-2\gamma^2}h^x \sum_j \left( \hat{\sigma}_j^+ e^{-i\theta_j(t)} + \hat{\sigma}_j^- e^{i\theta_j(t)} \right)
- \frac{g^2}{\omega_0}
\end{split}
\end{equation}
where $\theta_j(t) = -4\gamma \operatorname{Im}\{\alpha_j(t)\}$, and the expectation value $\langle\cdot\rangle$ is taken on time evolved spin state $|\psi(t)\rangle$. Phonons act back on spins via the time-dependent displacement $\alpha_j(t)$, while transverse spin terms are dressed by time-dependent phase factors 
$\pm\theta_i(t)$ due to the coherent phonon displacements, and the phonon occupation number $n_i(t)=|\alpha_i(t)|^2$ follows the spin rotation.

The considered ansatz captures essential spin-phonon dynamics in the fast-phonon and weak-coupling regimes. However, its validity is restricted to situations where phonon states have negligible entanglement \cite{Stojanovi2008}. The ansatz breaks down when spin-phonon coupling becomes strong, necessitating a full quantum treatment of the phonons. 

In the following sections, we study the dynamics of the initial string with the Matrix Product State formalism for spin-phonon system allowing capturing the system's dynamics in low phonon frequencies and strong coupling regimes, where the non-adiabatic effects and spin-phonon correlations become important.
}
\section{Initial state of the system}\label{sec:Observables}
As the initial state, we consider the product state of spins, with the leftmost $l$ spins pointing up, followed by $w$ spins pointing down, eventually containing $L - (l+w)$ spin pointing up, see Fig.~\ref{fig:initial_state} top panel. 
String length is equivalent to $w$. In the presence of phonons, the initial state is formally expressed as 
\begin{equation}
\begin{split}
\ket{\varPsi_{\rm ini}} = \prod_{j = 1}^l\ket{\up_j, n_j} \otimes \prod_{j = l+1}^{l+w}\ket{\dn_{j},n_j} \otimes \prod_{j = l+w+1}^{L}\ket{\up_{j},n_j}\
\end{split}\label{eq:initial_state},
\end{equation}
i.e., all spin states from the leftmost site $j = 1$ up to the site $j = l$ and from $j = l+w+1$ to the right end $j = L$ are in $\ket{\up}$. The phonon number state at location $j$ is denoted as $n_j$, 
{the eigenvalue of the operator}
$\hat{n}_j = \hat{a}^\dagger_j\hat{a}_j$.

Our goal is to provide a quantitative analysis of the string-breaking dynamics in the presence of phonons during the time evolution
\begin{equation}
    |\varPsi(t)\rangle = e^{-i t \hat{H}}|\varPsi_{\rm ini}\rangle,
\end{equation}
 {in the regime where the semi-classical description is invalid.} 
All expectation values $\langle \cdot\rangle$ are taken in the time evolved {state} 
$|\varPsi(t)\rangle$.

\begin{figure}[t!]
\includegraphics[width = 0.98\columnwidth]{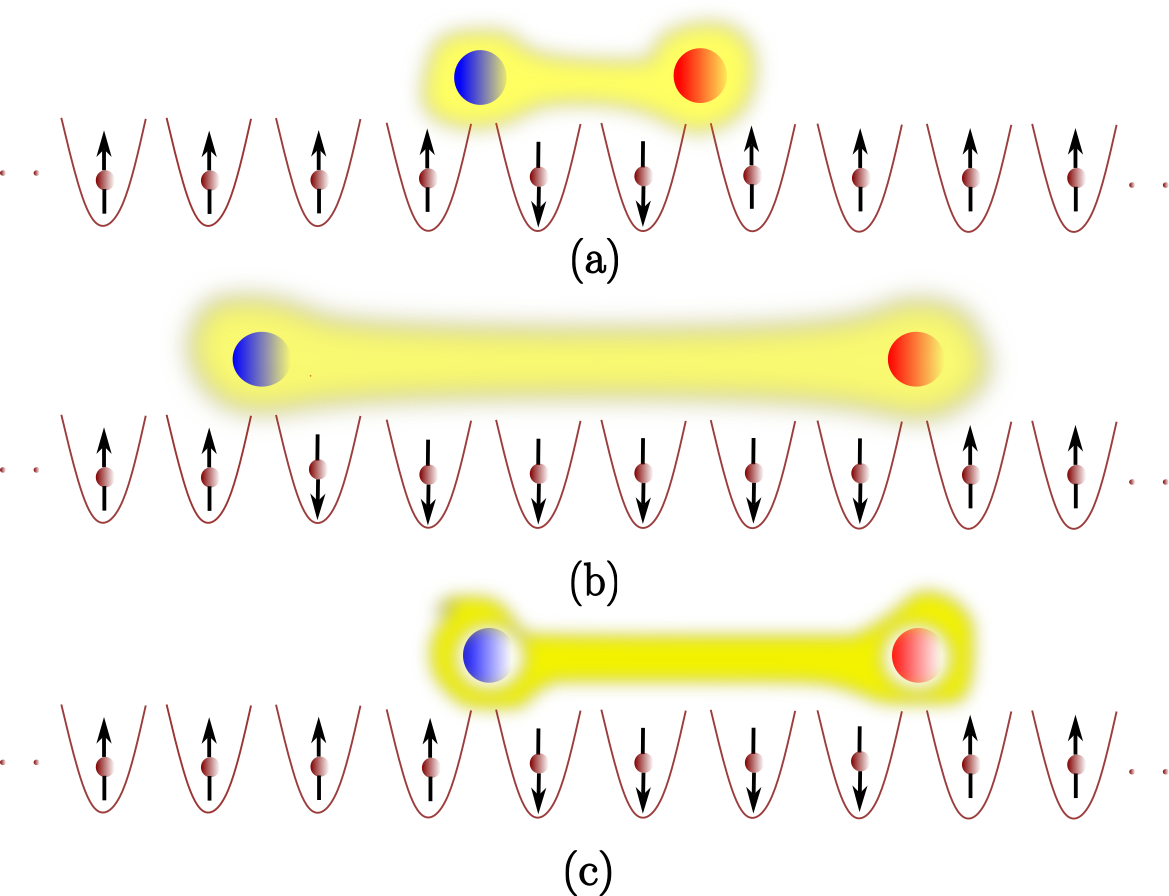}
\caption{Exemplary states representing schematically: (a) string-contraction, (b) string-expansion, and (c) displaced undistorted string in the same chain as in Fig.~\ref{fig:initial_state}.}
\label{fig:other_states}
\end{figure}

In the absence of the spin-phonon coupling ($g=0$), $n_j = 0$, the presence of magnetic transverse field $h^x$ breaks the string, and additional two domain walls are formed in the middle of the original string \cite{markus2020realtime, surace2021phdthesis}, see Fig.~\ref{fig:initial_state} bottom panel.
Stronger $h^x$ results in a faster string breaking, compare Fig.~7 in  Ref.~\cite{markus2020realtime}. The example shown in Fig.~\ref{fig:initial_state} bottom panel is a special case of string breaking where the lengths of broken strings are the same, the central $w = 4$ string $\ket{\downarrow \downarrow\downarrow\downarrow}$ maps to $\ket{\downarrow \uparrow \uparrow\downarrow}$. But there are two other possibilities where the lengths of strings are different: $\ket{\downarrow \downarrow\downarrow\downarrow}$ maps to $\ket{\downarrow \downarrow\uparrow\downarrow}$ or $\ket{\downarrow \uparrow \downarrow\downarrow}$. The number of different string-breaking configurations will grow with the increase of $w$. In addition to string breaking, there are other three possible mechanisms as visualized in Fig.~\ref{fig:other_states}: string contraction, string expansion, and displacement of the string without changing its length. They conserve the number of domain walls, but result in coordinate change of the domain walls. Figure \ref{fig:other_states}{\color{blue}(a)} shows for $w = 4$ only one string-contracted configuration $\ket{\uparrow \downarrow\downarrow\uparrow}$, other possible configurations are $\ket{\uparrow \downarrow\downarrow\downarrow}$, $\ket{\uparrow \uparrow\uparrow\downarrow}$ etc. Starting from an initial state, in general, the quantum evolution under Ising Hamiltonian is expected to result in a linear superposition of all outcome states from these mechanisms.

The average number of domain walls at Bond-$j$ between lattice sites $j$ and $j+1$ is defined as
 \begin{align}
 D_{j}(t) = \langle \hat{D}_j \rangle,~ \hat{D}_j = \frac{1}{2} - \frac{1}{2} \hat{\sigma}^z_j \hat{\sigma}^z_{j+1} 
 \label{eq:local_dw}
 \end{align}
measuring a nearest-neighbor correlation function in the $z$-spin component. 
To characterize the string-breaking mechanism we consider the average number of domain walls in two regions: (i) inside the initial string,
\begin{align}
 &D_{\rm in}(t) = \sum_{j = l+1}^{l+w-1} D_j(t),
\end{align}
and (ii) at the boundaries where the initial domain walls existed
\begin{align}
 &D_{\rm bd}(t) = D_{l}(t) + D_{l+w}(t)\;.
\end{align}
In principle, in the initial product state, the total number of domain walls inside the bulk of the system should change by even numbers only as a single spin-flip always changes the domain wall number by two. The example presented in top panel of Fig.~\ref{fig:initial_state} the $D_{\rm in}(t = 0) = 0$ and $D_{\rm bd}(t = 0) = 2$. Therefore string breaking would be defined as the time $t = \tau$ when $D_{\rm in} = 2$ for the first time, compare bottom panel Fig.~\ref{fig:initial_state} where $D_{\rm in} = 2$, $D_{\rm bd} = 2$. However, when the initial state is evolved under the Hamiltonian [Eq.~\eqref{eq:H_model}], the number of domain walls can change by ``any real value $\leq 2$''. Therefore one way to define the String-Breaking Time (SBT) $\tau$ which fulfills the following two criteria simultaneously: $D_{\rm in}(\tau)  \geq  D_{\rm bd}(\tau)$, 
and $D_{\rm in}(0<t<\tau)  <  D_{\rm bd}(0<t<\tau)$.

\begin{figure}[ht]
\centering 
\includegraphics[width=\columnwidth]{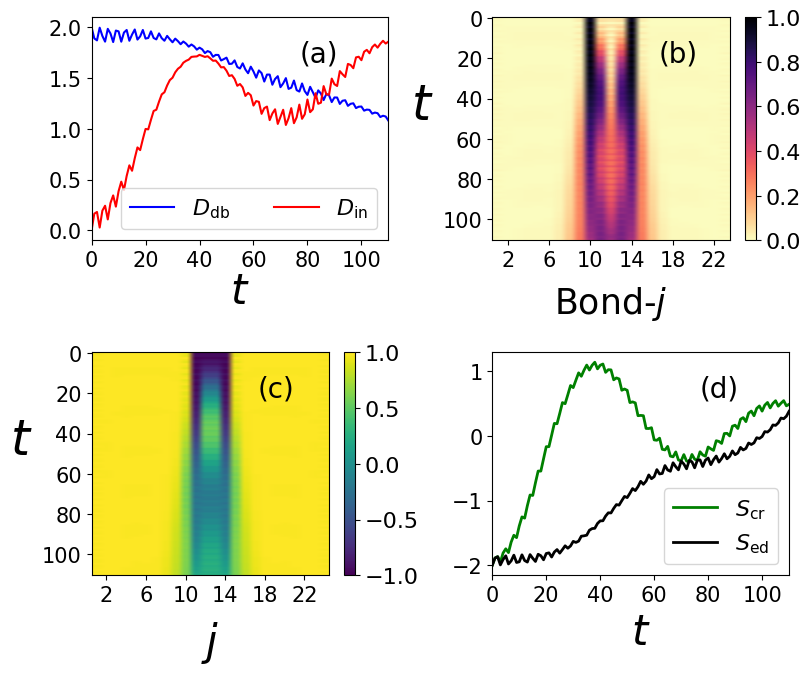}
\caption{(a) Time dynamics for $D_{j}$[Eq.~\eqref{eq:local_dw}] without spin-phonon coupling, (b) comparison of average domain walls inside the initial string = $D_{\rm in}$ and at its boundary = $D_{\rm bd}$, (c) longitudinal magnetization $\langle \hat{\sigma}^z_j \rangle$ as a function of time, (d) dynamics of magnetization [calculated according to Eq.~\eqref{eq:SBT_mag_measure}] at the edges $S_{\rm ed}$ and core $S_{\rm cr}$ of the initial string. System size $L = 24$ with open boundary condition, initial string width $w = 4$. $h^z = 1$, $h^x = 0.2$. 
}\label{fig:string_nophonon}
\end{figure}

According to this definition, Fig.~\ref{fig:string_nophonon} implies $\tau \approx 80$ where the exact crossing between two curves $D_{\rm in}$ and $D_{\rm bd}$ occurred, however   around $t \approx 40$ {these two curves are 
almost touching, suggesting also the string-breaking moment. To capture this kind of behavior, we provide a controllable measure allowing for identification string-breaking time based on expectation values and standard deviation of $D_{\rm bd}$, and $D_{\rm in}$.}
We consider the  { standard deviation} in the domain wall measurements  
\begin{align}
 &\Delta_{\rm in}(t) = \left[\langle \hat{D}^2_{\rm in} \rangle - \langle \hat{D}_{\rm in} \rangle^2\right]^\frac{1}{2} \notag\\
 &= \left[\sum_{i,j = l+1}^{l+w-1} \langle\hat{D}_i\hat{D}_j\rangle - \langle\hat{D}_i\rangle \langle\hat{D}_j\rangle \right]^\frac{1}{2} \notag\\
 &= \frac{1}{2} \left[\sum_{i,j = l+1}^{l+w-1} \langle\hat{\sigma}^z_i \hat{\sigma}^z_{i+1} \hat{\sigma}^z_j \hat{\sigma}^z_{j+1} \rangle - \langle\hat{\sigma}^z_i \hat{\sigma}^z_{i+1} \rangle \langle \hat{\sigma}^z_j \hat{\sigma}^z_{j+1} \rangle \right]^\frac{1}{2}
 \label{eq:in_devia}
\end{align}
which involves four-body correlation functions in the $z$-spin component.
A similar definition follows for the standard deviation $\Delta_{\rm bd}(t)$ at the boundaries where the initial domain wall existed. 
\begin{align}
 &\Delta_{\rm bd}(t) = \frac{1}{2} \Big[2 + 2\langle\hat{\sigma}^z_l \hat{\sigma}^z_{l+1} \hat{\sigma}^z_{l+w} \hat{\sigma}^z_{l+w+1} \rangle  -  \langle\hat{\sigma}^z_l \hat{\sigma}^z_{l+1} \rangle^2\notag\\&
 - 2\langle\hat{\sigma}^z_l \hat{\sigma}^z_{l+1} \rangle \langle \hat{\sigma}^z_{l+w} \hat{\sigma}^z_{l+w+1} \rangle 
 - \langle\hat{\sigma}^z_{l+w} \hat{\sigma}^z_{l+w+1} \rangle^2 \Big]^\frac{1}{2}\;.
 \label{eq:bd_devia}
\end{align}

Therefore we define the String-Breaking Time (SBT) $\tau$ which fulfills the following two criteria simultaneously
\begin{align}
&D^+_{\rm in}(\tau)  \geq  D^-_{\rm bd}(\tau), \notag\\
&D^+_{\rm in}(0<t<\tau)  <  D^-_{\rm bd}(0<t<\tau)\;.
\label{eq:SBT_measure}
 \end{align}
For convenience we used the short notations $D^\pm_{\rm in} = D_{\rm in} \pm \lambda \Delta_{\rm in}$ and $D^\pm_{\rm bd} = D_{\rm bd} \pm \lambda \Delta_{\rm bd}$.
The parameter $\lambda$ limits the error or controls the confidence level in the estimation of $\tau$. While the value of $\lambda$ is a matter of the investigator's preference, it is expected that the choice of larger $\lambda$ will result in shorter string-breaking time $\tau$.
In the next section, we will report the results for the choice $\lambda = 0.25$. 
It is also possible that for a longer time $t \gg \tau$ the number of domain walls inside $D^\pm_{\rm in}$  becomes smaller than $D^\pm_{\rm bd}$, and for a further longer time the system again satisfies $D^\pm_{\rm in} \geq  D^\pm_{\rm bd}$. To avoid confusion we stress that SBT  $\tau$ is the first time when the condition $D^+_{\rm in} \geq  D^-_{\rm bd}$   {  is met}.  {  In the light of the above, we consider string breaking at time $t \approx 40$.}

The time dynamics do not necessarily lead to pure string-like states. 
In particular, {a superposition of string-contracted and initial string-unbroken configurations} and {a superposition of string-breaking and initial string-unbroken configurations} are hardly distinguishable based only on the number of domain wall comparison \eqref{eq:SBT_measure}. 
In Section~\ref{sec:Results} we will see the string-expansion or string-displacement---which corresponds to a creation of extra domain walls outside of the initial string---appear only negligibly at the short time of evolution. Therefore, we will be concerned about the string-breaking and string-contraction, and will look for a way to distinguish them. 

The local magnetization (spin profile) as a function of time distinguishes the string-breaking from string-contraction, and it complements the string-breaking criterion based on the domain walls \eqref{eq:SBT_measure}. To be definite, we will extract the following two measures from the spin profile: total magnetization at the core of the string $S_{\rm cr}$, and at the edges of the string $S_{\rm ed}$, defined as
\begin{align}
 S_{\rm ed} = \langle \hat{\sigma}^z_{l+1} \rangle +  \langle \hat{\sigma}^z_{l+w} \rangle,~
 S_{\rm cr} = \sum_{j = l+2}^{l+w-1} \langle \hat{\sigma}^z_{j} \rangle\;.
\label{eq:SBT_mag_measure}
 \end{align}
If the initial state (top panel of Fig.~\ref{fig:initial_state}) for $w$ = 4 fully transforms into a string-breaking state, the pair $( S_{\rm cr}, S_{\rm ed}) = (-2, -2)$ maps to $(2, -2)$ or $(0, -2)$. On the other hand, if the initial state completely transforms into a string-contracted state, the pair $(S_{\rm cr}, S_{\rm ed}) = (-2, -2)$ maps to any pair from the set $\{(-2, 2), (-2, 0), (0,0), (2, 0), (0, 2)\}$. Figure \ref{fig:string_nophonon}{\color{blue}(c)-(d)} confirms the change in spin orientations at the core of the initial string at $t \approx 40$, which matches more with string-breaking configuration rather than string-contraction. Therefore, in the absence of phonons, the measured time $\tau$ according to \eqref{eq:SBT_measure} indeed corresponds to string-breaking. We will see below that this is not always the case.

\section{Results}\label{sec:Results}

{We numerically study the dynamics resulting from the Hamiltonian Eq.~\eqref{eq:H_model}  with the help of the time-dependent variational principle (TDVP) technique representing the wavefunction as a 
Matrix Product state}. We use the two-site time evolution scheme
as described in Ref.~\cite{PhysRevB.94.165116,PhysRevLett.107.070601}. 
For time evolution we use ITensor Julia library \cite{10.21468/SciPostPhysCodeb.4, 10.21468/SciPostPhysCodeb.4-r0.3}.

We introduce the cut-off for the maximal number of phonons allowed at any lattice site $j$ equal to $n_{\rm{max}}$.  {The value of this cutoff is taken to be much higher than the { average} number of phonons { $\langle \hat{n}_j \rangle$} at each site. In fact, we take $n_{\rm{max}}$ greater than the maximal ${ \langle \hat{n}_j \rangle}+2\Delta_j$  where $\Delta_j$ is the standard deviation of $n_j$. This {quite conservative} criterion assures that our results are converged with respect to $n_{\rm{max}}$. This is confirmed further by comparing the evolution for different $n_{\rm{max}}$ values.} 

We consider $L = 24$ spins in the open boundary conditions geometry, and the initial string width is set to $w = 4$. Throughout our study, we set $h^z = 1$ { (otherwise we will mention it explicitly)}. The vacuum/ground state is set as the initial state of the phonons. Interestingly for this parameter choice, the two configurations in Fig.~\ref{fig:initial_state}: initial state (top) and string-broken state (bottom) have the same energy expectation value for the spin part of the Hamiltonian $\langle \hat{H}_0 \rangle$. {Therefore in the absence of phonon this string-breaking is nothing but a resonant transition induced by the transverse magnetic field $h^x$.}

In the following sections, we present studies on string-breaking dynamics for both shallow, and deep trapping wells, in the case of weak, intermediate, and strong spin-phonon couplings.

\subsection{Shallow trapping wells, weak interaction}

We consider first shallow trapping potential wells setting $\omega_0  = 0.2$, and weak interaction regime $g \ll 1$.

\begin{figure}[t]
\centering 
\includegraphics[width=\columnwidth]{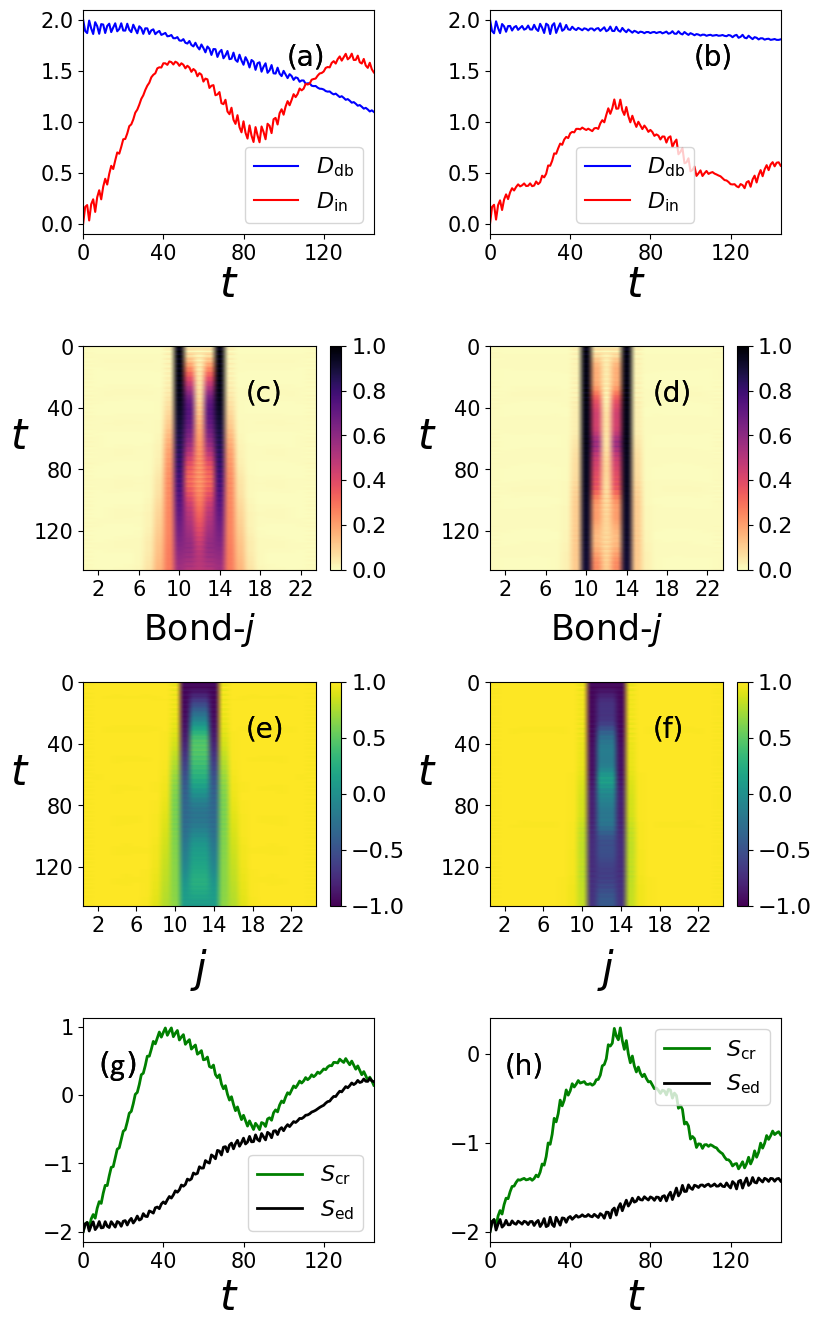}
\caption{Time dynamics in presence of spin-phonon couplings for  $g = 0.04$ (left column) and
$g = 0.08$ (right column). Panels (a) and (b) show dynamics of $D_{j}$[Eq.~\eqref{eq:local_dw}], (c)--(d) comparison of domain walls inside the initial string = $D_{\rm in}$ and at its boundary = $D_{\rm bd}$, (e)--(f) local magnetization as a function of time, (g)--(h) magnetizations at the core and edges of the initial string.  $\omega_0 = 0.2$, $n_{\rm max} = 4$. Other parameters are the same as in Fig.~\ref{fig:string_nophonon}. 
}\label{fig:string_1}
\end{figure}

\begin{figure}
\centering 
\includegraphics[width=\columnwidth]{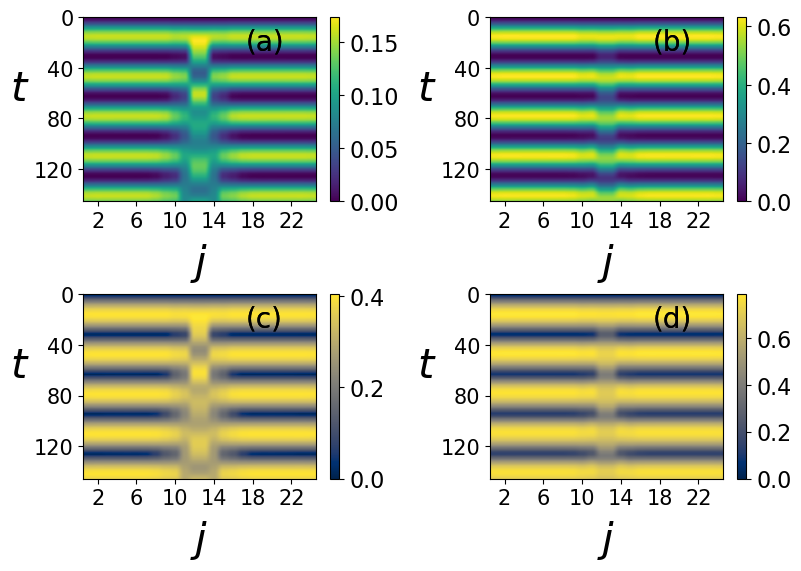}
\caption{ Time dynamics in presence of spin-phonon couplings for $g = 0.04$ (left column) and
$g = 0.08$ (right column). Panels (a) and (b) show dynamics of average phonon number $\langle\hat{n}_{j}\rangle$, (c) and (d) correspond to the standard deviation of the phonon number. 
Other parameters are the same as in Fig.~\ref{fig:string_1}. 
}\label{fig:string_1_phonon}
\end{figure}

\begin{figure}[h]
\centering 
\includegraphics[width=\columnwidth]{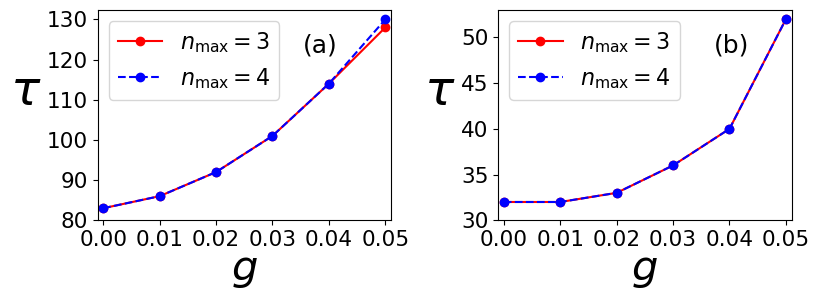}
\caption{String-breaking time (SBT), denoted as $\tau$, vs $g$ for $\omega_0 = 0.2$ and maximum allowed phonon number per site $n_{\rm max} = 3, 4$ for (a) $\lambda = 0$ and (b) $\lambda = 0.25$ during the SBT measurement \eqref{eq:SBT_measure}. 
All other parameters are the same as in Fig.~\ref{fig:string_1}.}
\label{fig:sbt_g_weak}
\end{figure}

Figure \ref{fig:string_1} shows domain wall dynamics at spin-phonon interaction strengths $g = 0.04, 0.08$ (we set the cutoff $n_{\rm{max}} = 4$). Comparing Figs.~\ref{fig:string_nophonon} and \ref{fig:string_1} we see that increasing $g$ increases the gap between the average $D_{\rm bd}$ and the first maximum of average $D_{\rm in}$. {At smaller $g = 0, 0.04$ the close encounter between  $D_{\rm bd}$ and $D_{\rm in}$ is observed after a short time of evolution, also around the same time the spin magnetization profiles evolve to spin-up states at the core, which is the sign of string-breaking---compare Fig.~\ref{fig:string_1} with Fig.~\ref{fig:initial_state}. For larger $g = 0.08$, up to the time-scale of study, the gap between $D_{\rm bd}$ and $D_{\rm in}$ is so large that it is unreasonable to consider any string-breaking.

Interaction excites the phonons in the system
as shown in Fig.~\ref{fig:string_1_phonon} while initially, no phonon exists.
The phonon number outside the initial string maintains its oscillation with time having a frequency of approximately $2\pi/\omega_0 = 2\pi/0.2 = 31$. But inside the initial string region where the dynamics significantly change the spin states, we see distortions of the oscillating phonon number profiles.

Figure \ref{fig:sbt_g_weak} shows the increase of string breaking time upon increasing spin-phonon coupling parameter $g$ at $\omega_0 = 0.2$. Therefore the presence of the phonon enhances the string stabilization  
\footnote{As described in the definition of the string-breaking time, Eq.~\eqref{eq:SBT_measure}, larger error $\lambda = 0.25$ reduces the time $\tau$ compared to the $\lambda = 0$ case. We assume  $\lambda = 0.25$ from now on (the results do not change significantly for $\lambda \in [0.2,0.3]$).  }.

\subsection{Shallow trapping wells, strong interaction}

For stronger spin-phonon interactions $g$ the number of generated phonons increases. While keeping the same $\omega_0 = 0.2$, we observe a significant difference in the dynamics.
As shown below, the string-contraction dominates over string-breaking in that case.

\begin{figure}[t!]
\centering 
\includegraphics[width=\columnwidth]{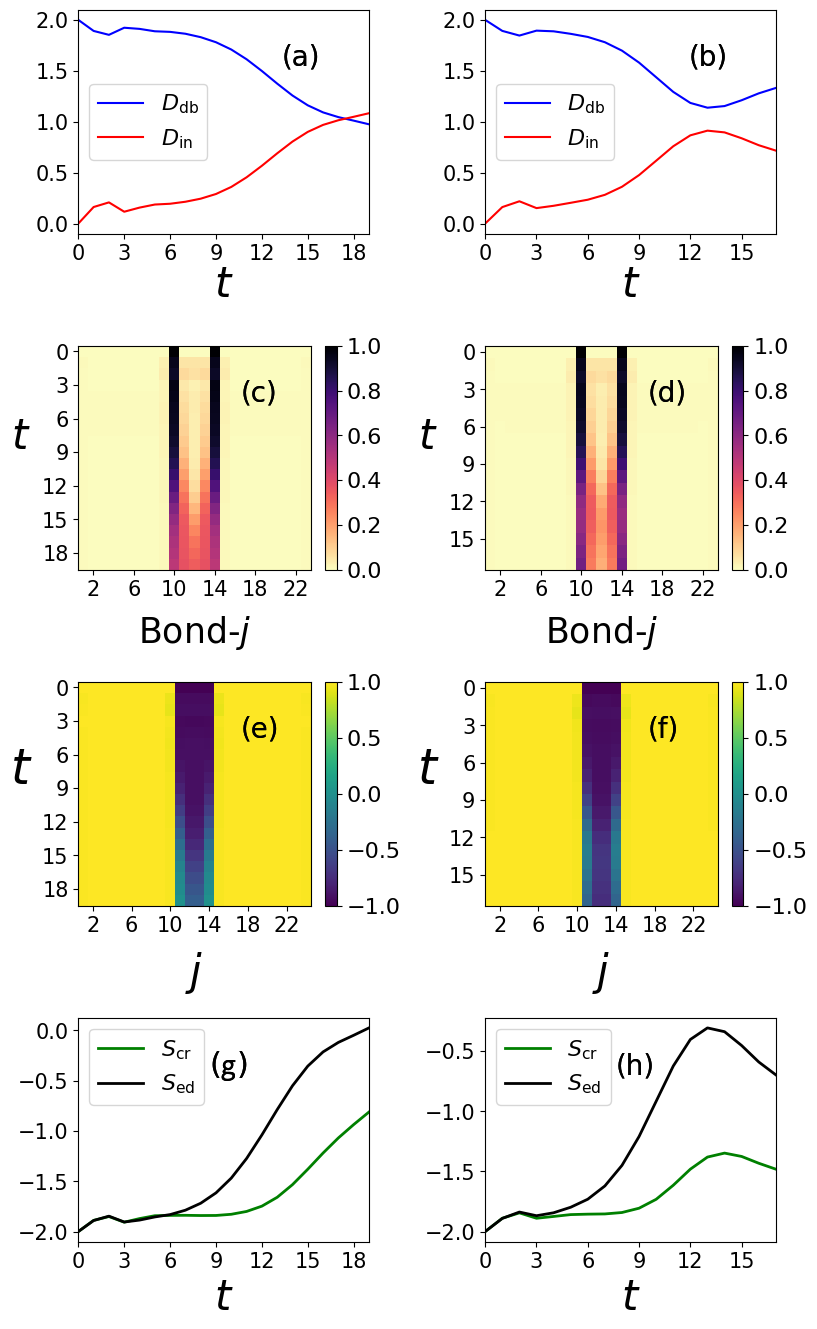}
\caption{Time dynamics in the presence of strong spin-phonon coupling for $g = 0.23$ (left column) and $g = 0.28$ (right column). Panels (a) and (b) show dynamics of $D_{j}$, (c)--(d) comparison of domain walls inside the initial string = $D_{\rm in}$ and at its boundary = $D_{\rm bd}$, (e)--(f) local magnetization as a function of time, (g)--(h) magnetizations at the core and edges of the initial strings.  $\omega_0 = 0.2$, $n_{\rm max} = 20$. Other parameters are the same as in Fig.~\ref{fig:string_nophonon}. 
}\label{fig:string_2}
\end{figure}
 Figure \ref{fig:string_2} shows the domain wall and spin profile dynamics as functions of time for interaction strength $g = 0.23, 0.28$. The domain walls $D_{\rm in}, D_{\rm bd}$ cross or nearly touch at a certain time associated with string-breaking or contraction. The spin profiles imply the flipping of spin-down states (with a certain probability) at the edges of the initial string while spins at the core remain more negatively magnetized compared to the edges. That suggests the string-contraction prevails over other possible mechanisms e.g.~string-breaking---compare $S_{\rm cr}$ (green) and $S_{\rm ed}$ (black) curves in Fig.~\ref{fig:string_1} with those in Fig.~\ref{fig:string_2}.

\begin{figure}[h]
\centering 
\includegraphics[width=0.9\columnwidth]{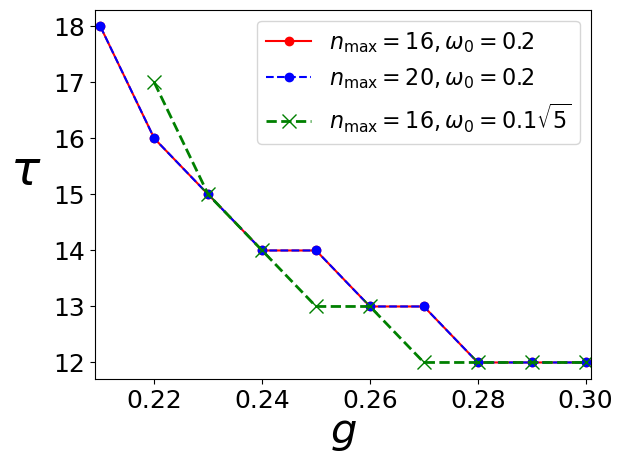}
\caption{String-contraction time vs $g$ for $\omega_0 = 0.2$ 
(blue circles connected by a line)
at stronger interaction strength $g$, and the maximum allowed phonon number per site $n_{\rm max} = 16, 20$ for $\lambda = 0.25$ during the SBT measurement \eqref{eq:SBT_measure}. 
Green crosses connected by a dashed line are for an incommensurate $\omega_0 = 0.1\sqrt{5}$.
All other parameters are the same as in Fig.~\ref{fig:string_1}.
}
\label{fig:sbt_g_strong}
\end{figure}
Although we aim to discuss the string-breaking in this work, it is interesting to see how this string-contraction mechanism changes with spin-phonon interaction. The definition of string-breaking time at Eq.~\eqref{eq:SBT_measure} also works for finding the string-contraction time. Figure \ref{fig:sbt_g_strong} shows decreasing string-contraction time $\tau$ with interaction strength $g$. It may be tempting to associate such opposite behavior with the possible resonance between $\hat{H}_{\rm ph} = \omega_0 \sum_j\hat{a}^\dagger_j\hat{a}_j $ and the longitudinal magnetic potential term in $\hat{H}_0$ which {might happen}  because of the presence of allowed phonon numbers {$n_j$ = }$5, 10, 15$ when $n_{\rm max} \geq 15$. But a slight change of $\omega_0$ from $0.2$ to $0.1\sqrt{5}$ does not alter the qualitative behavior of the decreasing $\tau$ with increasing $g$ 
as shown in Fig.~\ref{fig:sbt_g_strong}.

\subsection{Shallow trapping wells, intermediate interaction}

\begin{figure}[h]
\centering 
\includegraphics[width=0.95\columnwidth]{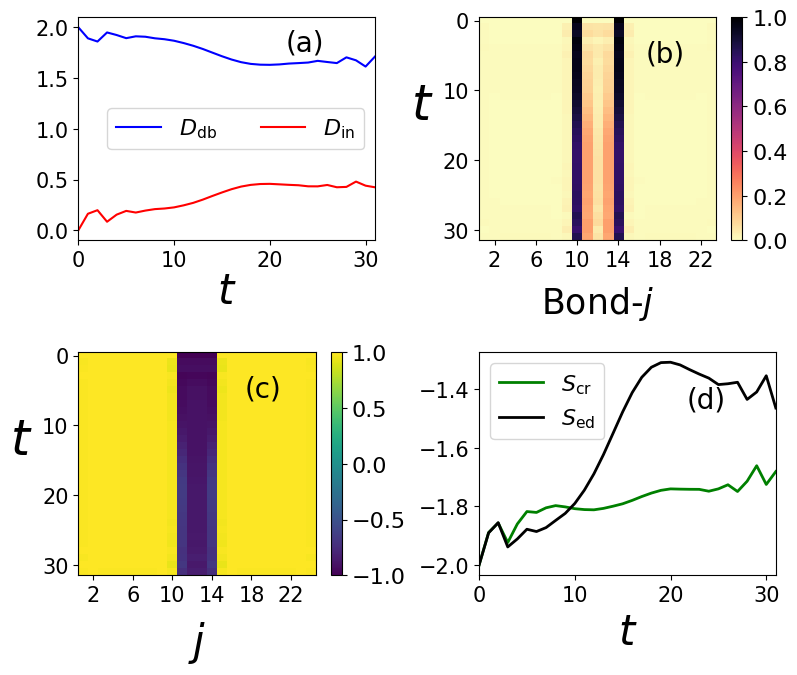}
\caption{(a) Time dynamics for $D_{j}$[Eq.~\eqref{eq:local_dw}], (b) comparison of average domain walls inside the initial string = $D_{\rm in}$ and at its boundary = $D_{\rm bd}$, (c) longitudinal magnetization $\langle \hat{\sigma}^z_j \rangle$ as a function of time, (d) dynamics of magnetization [calculated according to Eq.~\eqref{eq:SBT_mag_measure}] at the edges $S_{\rm ed}$ and core $S_{\rm cr}$ of the initial string. $g = 0.18$, $\omega_0 = 0.2$, $n_{\rm max} = 14$, $h^z = 1$, $h^x = 0.2$. 
}\label{fig:string_inter_3}
\end{figure}

{At the intermediate interaction regime: $0.08 < g < 0.21$ no significant sign of either string-breaking or string-contraction is observed up to the time scales of numerical runs, as domain wall curves $D_{\rm in}$ and $D_{\rm bd}$ always maintain a large gap. The results are shown in 
Fig.~\ref{fig:string_inter_3} 
for $g = 0.18$ with $n_{\rm max} =14$. {The initial domain walls remain relatively stable despite the coupling to phonons as indicated by $D_{\rm db}(t)\approx 2$. The dynamics of $D_{\rm in}$ indicates that many possible intermediate quantum states participate in it as it is also signaled by 
magnetization profiles.}

\subsection{Deep trapping wells}
\begin{figure}[ht]
\centering 
 \includegraphics[width=0.9\columnwidth]{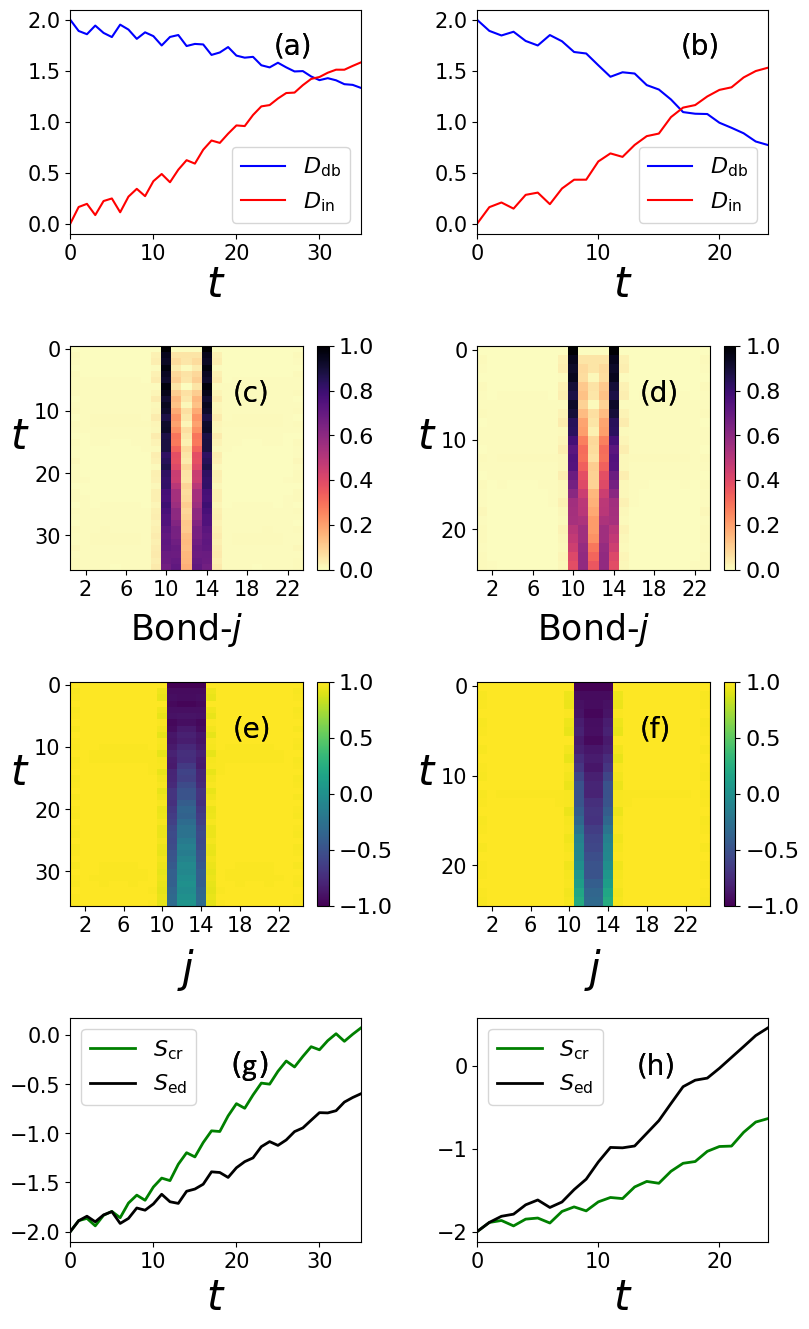}
\caption{The dynamics for $\omega_0 = 1$ and $g = 0.2$ (left column) and $g= 0.3$ (right column). $n_{\rm max} = 8$, $\lambda = 0.25$ during the SBT measurement \eqref{eq:SBT_measure}. All other parameters are the same as in Fig.~\ref{fig:string_1}. (a)--(b) Comparison of domain walls inside the initial string = $D_{\rm in}$ and at its boundary = $D_{\rm bd}$, (c)--(d) Domain wall dynamics, (e)--(f) longitudinal magnetization dynamics, (g)--(h) time evolution of magnetizations at the edges and core of the initial string.}
\label{fig:string_omg1}
\end{figure}

Consider now deep trapping potentials at each lattice site. Let us set $\omega_0 = 1$. The domain wall and local spin dynamics are shown in Fig.~\ref{fig:string_omg1}. From the magnetization profiles [Fig.~\ref{fig:string_omg1}{\color{blue}(e)-(h)}] increasing $g$ results in a transition from ``string-breaking'' to ``string-contraction'' regimes.

{The corresponding dynamics of phonon numbers $n_j$ and their standard deviations are shown in Fig.~\ref{fig:string_omg1_phonon}. Observe significant photonic excitations occurring at the positions of the original domain walls. For the time range studied the energy is transferred from the spins to the (initially empty) photonic modes at the walls while at other positions weak oscillations with frequency $\omega_0$ are observed only. The number of generated phonons is much smaller than for shallow wells for the same $g$ (as expected since each phonon carries now a bigger energy). }

\begin{figure} 
\centering 
 \subfigure{\includegraphics[width=0.49\columnwidth]{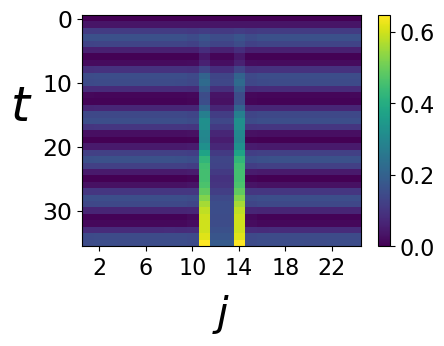}\llap{
  \parbox[b]{0.24\textwidth}{{\bf(a)}\\\rule{0ex}{1.25in}}}}
\subfigure{\includegraphics[width=0.49\columnwidth]{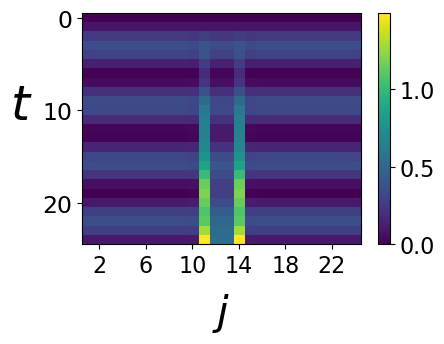}\llap{
  \parbox[b]{0.24\textwidth}{{\bf(b)}\\\rule{0ex}{1.25in}}}}\\
  \subfigure{\includegraphics[width=0.49\columnwidth]{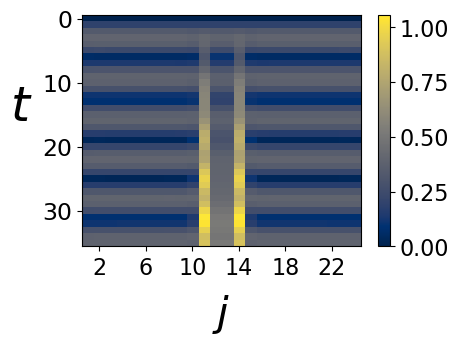}\llap{
 \parbox[b]{0.24\textwidth}{{\bf(c)}\\\rule{0ex}{1.25in}}}}
\subfigure{\includegraphics[width=0.49\columnwidth]{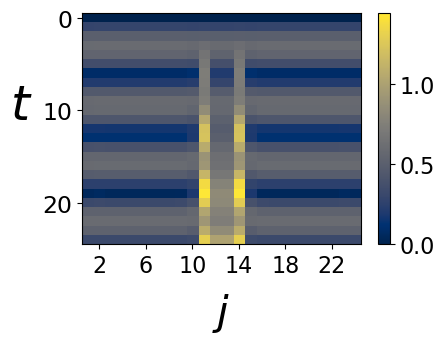}\llap{ 
 \parbox[b]{0.24\textwidth}{{\bf(d)}\\\rule{0ex}{1.25in}}}}

\caption{The phonon dynamics for results in Fig.~\ref{fig:string_omg1}. Left [(a), (c)] and right panels [(b), (d)] are at $g = 0.2$ and $g = 0.3$, respectively. The first row [(a), (b)] shows the average phonon number while the second row [(c), (d)] shows its standard deviation. All the results are at $n_{\rm max} = 8$.}
\label{fig:string_omg1_phonon}
\end{figure}

\begin{figure} 
\centering 
\includegraphics[width=0.85\columnwidth]{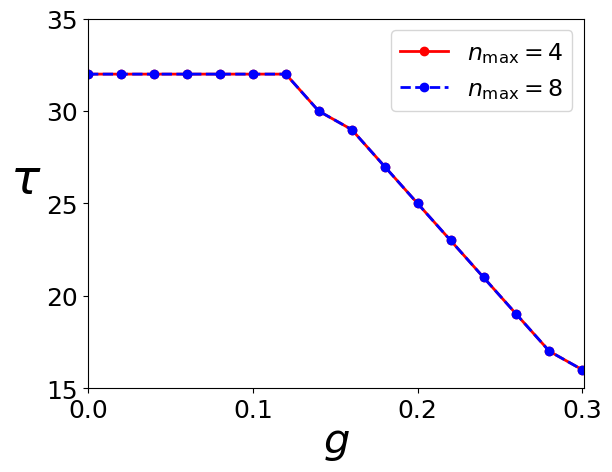}
\caption{String-breaking/contraction time as a function of $g$ for longer range of values at $\omega_0 = 1$. Other setups are the same as in Fig.~\ref{fig:string_omg1}. The results for $n_{\rm max} = 4, 8$ match well indicating the convergence w.r.t.~$n_{\rm max}$.}
\label{fig:sbt_g_omg1}
\end{figure}

The string-breaking or string-contraction times as a function of $g$ are shown in Fig.~\ref{fig:sbt_g_omg1}. { While for shallow potential wells the weak coupling to phonons slowed down the dynamics (increasing the SBT) here, for $\omega_0=1$   the weak spin-phonon interactions $g \leq 0.1$  do not affect the string-breaking time $\tau$.} For stronger interactions, a monotonic decrease of  $\tau$ with $g$ is observed which, for $g>0.25$ corresponds to a faster {string-contraction.} 
  Flipping the orientation of the longitudinal magnetic field ($h^z = -1$) is expected to induce string-expansion instead of string-contraction at strong $g$---see Fig.~\ref{fig:string_negative_hz}.

\begin{figure}[t!]
\centering 
\subfigure{\includegraphics[width=0.49\columnwidth]{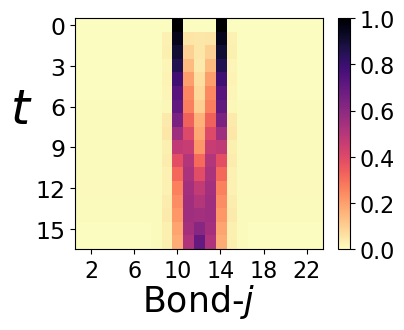}\llap{
\parbox[b]{0.32\textwidth}{{\bf(a)}\\\rule{0ex}{0.8in}}}}
\subfigure{\includegraphics[width=0.49\columnwidth]{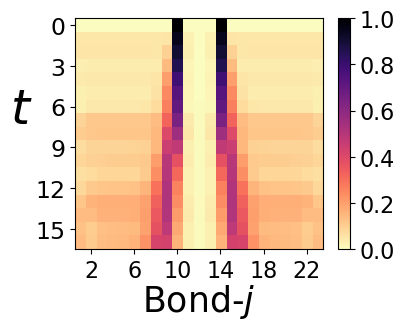}\llap{
\parbox[b]{0.32\textwidth}{{\bf(b)}\\\rule{0ex}{0.8in}}}}\\
\subfigure{\includegraphics[width=0.49\columnwidth]{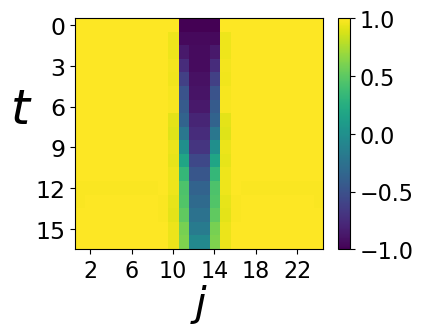}\llap{
\parbox[b]{0.32\textwidth}{{\bf(c)}\\\rule{0ex}{0.8in}}}}
\subfigure{\includegraphics[width=0.49\columnwidth]{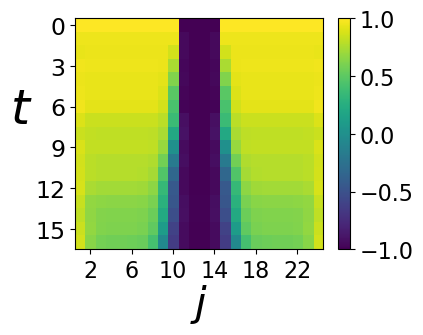}\llap{ 
\parbox[b]{0.32\textwidth}{{\bf(d)}\\\rule{0ex}{0.8in}}}}
\caption{{The left [(a), (c)] and right [(b), (d)] columns represent the dynamics for $h^z = 1$ and $h^z = -1$ respectively, at $\omega_0 = 1$, strong interaction regime $g = 0.6$. (a)--(b) Domain wall dynamics, and (c)--(d) local spin dynamics. All the results are at $n_{\rm max} = 8$ with closed boundary conditions (to avoid significant effects coming from open boundaries when $h^z = -1$).  String-contraction dominates the dynamics for $h^z = 1$, while string expansion occurs for $h^z = -1$.}}
\label{fig:string_negative_hz}
\end{figure}

\section{Discussion and Conclusions}\label{sec:Conclusions}

We have considered the dynamics of the initial string in a modified Ising chain. We assume that each site vibrates with a specific frequency.
The dynamics of the initial string is significantly modified by the coupling of sites to   { phononic} modes. Such a model mimics
a realistic quantum simulation platform, where each spin vibrates with respect to its mean position. To make the model as simple as possible we consider identical harmonic potentials for the trapping wells. 

The quantum Ising chain is known to capture the properties of particle creation-annihilation and dynamics of gauge-flux relevant for lattice-gauge-theories, e.g.~$\mathbb{Z}_2$ LGT. Initially prepared string eventually breaks. The presence of vibrations (and spin-phonon coupling) significantly modifies this simple picture. 
{  Due to phonon-dressing of the transverse field operator, in the context of spin-breaking mechanisms, the spin-phonon coupling results in reduced gauge-field induced interaction between (anti-)particles, thus supressing creation of particle-antiparticle pair.
}

We have considered different possible regimes. The first one corresponds to the case when the frequency of vibrations is much lower than the longitudinal magnetic field {strength}. Here the weak coupling to phonons seems to slow down the spin dynamics due to the creation of phonons. In effect the string breaking time increases with the spin-phonon coupling strength, $g$. A further increase of $g$ modifies significantly the dynamics, the domain walls seem to freeze, while at still larger $g$ one may identify other than string-breaking mechanisms. In particular, for a strong $g$ we observe the string contraction instead of the string-breaking. 

For larger vibration frequency, of the order of the transverse field a weak coupling to phonons practically does not affect the dynamics and the observed string breaking time is similar to that without phononic background. Beyond some critical $g$ value the observed characteristic time starts to decrease, and spin dynamics become faster. Depending on the sign of
the longitudinal field one observes either string expansion or string contraction. 

Our study of the spin-phonon coupling effect provides a realistic scenario in quantum simulation setups in tweezers or trapped ions physics.  Moreover, it could motivate the study of many-body confinement effects in mesoscopic or macroscopic condensed matter systems{, in the presence of environmental influence}. 

\acknowledgments
The work of A.M. and J.Z. was
funded by the National Science Centre, Poland, project
2021/03/Y/ST2/00186 within the QuantERA II Programme (DYNAMITE) that has received funding from the European
Union Horizon 2020 research and innovation programme
under Grant agreement No 101017733. We gratefully acknowledge Polish high-performance computing infrastructure PLGrid (HPC Center: ACK Cyfronet AGH) for providing computer facilities and support within computational grant no. PLG/2024/017289.
The research has been also supported by a grant from the Priority Research Area (DigiWorld)
under the Strategic Programme Excellence Initiative at Jagiellonian University (J.Z.).
ICFO-QOT group acknowledges support from: European Research Council AdG NOQIA; MCIN/AEI (PGC2018-0910.13039/501100011033, CEX2019-000910-S/10.13039/501100011033, Plan National FIDEUA PID2019-106901GB-I00, Plan National STAMEENA PID2022-139099NB, I00, project funded by MCIN/AEI/10.13039/501100011033 and by the “European Union NextGenerationEU/PRTR" (PRTR-C17.I1), FPI); QUANTERA DYNAMITE PCI2022-132919, QuantERA II Programme co-funded by European Union’s Horizon 2020 program under Grant Agreement No 101017733; Ministry for Digital Transformation and of Civil Service of the Spanish Government through the QUANTUM ENIA project call - Quantum Spain project, and by the European Union through the Recovery, Transformation and Resilience Plan - NextGenerationEU within the framework of the Digital Spain 2026 Agenda; Fundació Cellex; Fundació Mir-Puig; Generalitat de Catalunya (European Social Fund FEDER and CERCA program; Barcelona Supercomputing Center MareNostrum (FI-2023-3-0024); Funded by the European Union. Views and opinions expressed are however those of the author(s) only and do not necessarily reflect those of the European Union, European Commission, European Climate, Infrastructure and Environment Executive Agency (CINEA), or any other granting authority. Neither the European Union nor any granting authority can be held responsible for them (HORIZON-CL4-2022-QUANTUM-02-SGA PASQuanS2.1, 101113690, EU Horizon 2020 FET-OPEN OPTOlogic, Grant No 899794, QU-ATTO, 101168628), EU Horizon Europe Program (This project has received funding from the European Union’s Horizon Europe research and innovation program under grant agreement No 101080086 NeQSTGrant Agreement 101080086 — NeQST); ICFO Internal “QuantumGaudi” project.
\bibliography{biblio}{}

\end{document}